\documentclass[a4paper,12pt]{article}
\usepackage{geometry}
\usepackage{url}        
\usepackage{hyperref}   
\usepackage{CJKutf8}
\usepackage{titlesec}
\usepackage{fancyhdr}
\usepackage{lmodern}
\usepackage{setspace}
\usepackage{amsmath, amssymb}
\usepackage[numbers]{natbib}
\usepackage{indentfirst}
\usepackage{enumitem} 
\usepackage{pifont}
\usepackage{algorithm}
\usepackage{graphicx}
\usepackage{multirow}
\usepackage[noend]{algpseudocode}
\usepackage[table,xcdraw]{xcolor}
\usepackage[noend]{algpseudocode}
\usepackage{fancyhdr}  

\titleformat{\section}{\large\bfseries}{\thesection.}{1em}{}
\titleformat{\subsection}{\normalsize\bfseries}{\thesubsection.}{1em}{}

\title{\textbf{Knowledge Migration Framework for Smart Contract Vulnerability Detection}}
\author{
    Luqi Wang\textsuperscript{1}, Wenbao Jiang\textsuperscript{2} \\[1em]
    \small \textsuperscript{1}Beijing Information Science and Technology University, College of Computer Science, Beijing, China\\
}
\date{} 

\begin{document}
\begin{CJK}{UTF8}{gbsn}

\maketitle 

\begin{abstract}
As a cornerstone of blockchain technology in the 3.0 era, smart contracts play a pivotal role in the evolution of blockchain systems. Unfortunately, with the rapid growth of smart contract-related data, modern deep learning models often face challenges such as limited understanding of contract semantics and high computational overhead during training. Moreover, traditional detection techniques often rely on sensitive transaction information present in the initial smart contract, posing risks to blockchain privacy and security.

To address these limitations, an AF-STip smart contract vulnerability detection framework incorporating efficient knowledge migration is proposed. AF-STip employs the teacher network as the main model and migrates the knowledge processed by the smart contract to the student model using a data-free knowledge distillation method. The student model enhances its vulnerability detection capabilities, while reducing computational overhead. Additionally, an adaptive fusion module is proposed to further improve vulnerability feature extraction. Experimental results demonstrate that the STip model attains an average F1 detection score of 91.16\% for four vulnerabilities without disclosing smart contract data. To validate the lightweight migration approach, the student model achieves an accuracy of 91.02\% and an F1 score of 90.46\% on a novel vulnerability type. To the best of our knowledge, AF-STip is the first model to apply data-free knowledge migration to smart contract vulnerability detection, achieving exceptional performance with reduced computational overhead.
\end{abstract}


\section{Introduction}
With the development of the "Metaverse and Value Internet 3.0" era, the immutability, decentralization, and support for smart 
contracts in blockchain technology have driven the construction of the internal economic system of the metaverse. Blockchain originated from Bitcoin, 
using cryptography and chain storage to verify data, and enabling automatic execution of transactions through smart contracts without third-party intervention, 
thereby reducing transaction trust costs. However, the immutability of blockchain also introduces security risks for smart contracts.
\par The rapid development of blockchain platforms such as Ethereum has led to a notable increase in the deployment of smart contracts on modern blockchains.
 It is possible for developers to create and publish contracts without requiring any special permissions. As Mohanta et al. \cite{mohanta2018overview} have observed, the number of 
 new smart contracts on Ethereum may reach hundreds of thousands per month. Nevertheless, the considerable obstacles to entry in the field of smart contract 
 development require that developers not only master the intricacies of programming languages such as Solidity and Vyper, but also possess a comprehensive grasp 
 of the foundational tenets governing blockchain operations. Moreover, once deployed, smart contracts are immutable, and any subsequent updates incur a fee. 
 In the event that vulnerabilities are present, it is possible for hackers to rapidly exploit them for the purpose of launching attacks. In July 2023,
  a smart contract on Ethereum, written with a specific version of Vyper, was found to contain a reentrancy lock flaw \cite{inspex2023curveattack}, which resulted in the theft 
  of approximately \$\ 25 million worth of digital assets from the CRV/ETH pool. In the same year, a report from Certik \cite{certik2023web3security} revealed that Ethereum caused \$\ 686 
  million in losses. This emphasises the critical necessity for effective vulnerability detection in order to guarantee the secure operation of smart contracts on the blockchain.

\par To address the above issues, smart contract vulnerability detection is divided into two categories: one is the 
traditional detection tools based on static analysis and symbolic execution (e.g. Mythril, Securify, etc.). 
The other category is automated detection methods based on deep learning. Traditional detection tools usually focus on directly analysing the 
code structure, which usually requires costly computational resources and human intervention, and has limitations when dealing with complex contracts.
 Deep learning-based methods can improve the accuracy of vulnerability detection by data-driven analysis of contract code. Currently there are numerous 
 scholars have conducted a lot of research. xueli Shen \cite{shen2023smart}proposed a parallel hybrid model combining convolutional neural network (CNN) 
 and long-short-term memory network (LSTM) for feature extraction and using gating units to improve the data processing efficiency. gogineni \cite{gogineni2020multi} et al. 
 utilised the memory mechanism of LSTM and based on the opcodes of the historical contracts only for the training.Kalra \cite{kalra2018zeus} et al. proposed the ZEUS tool 
 in conjunction with static analysis to transform smart contracts into symbolic paths and classify them through deep learning. 
Each of these methods has its own advantages in terms of accuracy and performance, but generally suffers from the following drawbacks:
\begin{enumerate}[label=(\arabic*)]
  \item High computational overhead: Existing deep learning architectures such as CNN-LSTM and GRU-LSTM can capture rich features, but their large number of parameters and high computational requirements make them difficult to deploy in decentralised systems.
  \item High resource dependency: Existing models rely heavily on data quality and struggle to generalise well when faced with poorly labelled smart contract datasets.
\end{enumerate}
\par In conclusion, this paper proposes an effective adaptive fusion smart contract vulnerability detection model based on data-free distillation, which addresses the shortcomings of existing methodologies.The main contributions of this paper are as follows:
\begin{enumerate}[label=(\arabic*)]
  \item A teacher-student neural network model AF-STip based on data-free distillation was constructed. The model captures the semantic information of smart contract source code by introducing word embeddings to build the dataset. During the pretraining and fine-tuning stages, it applies a data-free knowledge transfer strategy to perform knowledge distillation on both the embedding vector encoder and the adaptive attention matrix of the fusion module. The student model is guided by the teacher's knowledge in the absence of additional training data, further improving the generalisation ability of the student model in downstream tasks. 
  \item A new adaptive fusion module is developed. The teacher model is capable of capturing fine-grained local features and long-range global dependencies while processing different semantic features, enhancing the collaborative modelling capability of feature interactions. Specifically, the module integrates multi-level features in parallel with weighted fusion, enhancing the global relationships between features. It effectively captures key vulnerability features while balancing the complex dependencies in the contract source code, improving the robustness of the model.
  \item The model proposed in this paper adopts a lightweight architecture. By combining data-free knowledge transfer with the design of the adaptive attention fusion module, the student model effectively inherits the semantic understanding capabilities of the teacher model, optimising computational efficiency while improving performance. Therefore, this model is not only applicable to the analysis of vulnerabilities in smart contracts, but can also be extended to other areas of text analysis, providing new insights for related fields.
  \item In order to promote the development of the field of smart contract vulnerability detection, this study chooses to disclose our processing dataset 
  and construction method. Published on https://github.com/Frances-Kay/AF-STip.git. This dataset contains various vulnerabilities such as timestamp 
  vulnerabilities, reentrancy vulnerabilities, 
  delegate call vulnerabilities and CDAV (Contract Deployment Address) vulnerabilities, all processed using embedding vector encoders. 
\end{enumerate}

\section{Related work}
\subsection{Knowledge distillation}
The concept of Knowledge Distillation (KD) was initially proposed by Hinton et al\cite{hinton2015distilling}. The method 
entails transferring knowledge from a teacher model to a student model, thereby reducing the computational 
overhead while maintaining high performance. The core principle involves softening the Softmax output by a 
temperature parameter to enhance the similarity between categories and improve the generalisation ability of 
the student model. The underlying formula is $\mathrm{q}_{\mathrm{i}}=\frac{\exp \left(\mathrm{z}_{\mathrm{i}} / \mathrm{T}\right)}{(\mathrm{Ej}) \exp \left(\mathrm{z}_{\mathrm{j}} / \mathrm{T}\right)}$, where $z_{i}$ represents the teacher model output and $q_{i}$  denotes the temperature-smoothed output. 
\par The research methods employed in knowledge distillation can be broadly classified into three 
categories. The initial category encompasses temperature-based distillation methodologies. 
In a related contribution, Wei Y. \cite{wei2024dynamic} put forth a dynamic temperature adjustment mechanism that entails
 a reduction in temperature over the course of training steps, with the objective of optimising the 
 student model. In a similar vein, Long J. \cite{long2024mkdat} adopted a dual-temperature mechanism, 
 utilising higher temperatures for those samples that are more challenging to classify and lower temperatures 
 for those that are less so. The second category comprises feature alignment and self-distillation methods. 
 The paper \cite{zagoruyko2016paying} employs mean squared error to align the probabilities of the teacher and student models, 
 whereas the Born-Again model \cite{furlanello2018born} utilises self-distillation for the transfer of knowledge. 
 The paper \cite{wang2023improving} improves the performance of the student model by regularising feature norms. Similarly, 
 Park et al. \cite{park2019relational} put forth a proposal for the transfer of knowledge via the establishment of 
 relationships between samples (such as distance or angle) with the objective of maintaining the diverse 
 features of the teacher model. Nevertheless, in high-dimensional data spaces, this method is vulnerable 
 to noise and may encounter difficulties in effectively capturing the intricate distribution of data. 
 Furthermore, knowledge distillation has been extensively employed in a multitude of domains. 
 The Cross-Task Distillation framework \cite{yang2022cross} is a multi-task learning method that enhances the performance 
 of the student model in multiple tasks by sharing teacher knowledge across tasks. These knowledge 
 distillation methods continue to rely on external labels or source training data, which constrains their 
 applicability in scenarios with no or limited labelled data. The data-free distillation method \cite{shao2022review} 
 addresses this limitation by enabling student model training guided by the teacher model's prior knowledge 
 without the need for source data. This approach shows promise for applications in privacy protection 
 and blockchain domains.

\subsection{Detecting Vulnerabilities in Smart Contracts}
The advent of blockchain technology in the 3.0 era has underscored the critical 
importance of smart contract vulnerability detection in the field of blockchain security. 
At present, smart contract vulnerabilities are typically classified according to their 
manifestation into categories such as contract-level vulnerabilities, arithmetic vulnerabilities, 
network-layer vulnerabilities, and external dependency vulnerabilities \cite{vidal2024vulnerability}. The majority of 
traditional smart contract vulnerability detection tools \cite{luu2016making,ruskin1980mythril,tsankov2018securify,mossberg2019manticore}, including Oyente and Mythril, rely 
on symbolic execution techniques. These tools examine contract code through control flow graphs 
or static analysis of bytecode. However, traditional detection tools often depend on expert-defined 
original vulnerability specifications to identify potential issues within contracts. As a result, they are 
limited in their 
ability to detect complex vulnerabilities and fail to address more sophisticated smart contracts.
\par In contrast, deep learning-based vulnerability detection methods \cite{zhang2022cbgru,liu2023smart,zhang2022novel,zhang2022spcbig} effectively address vulnerability 
detection in complex smart contracts through automated feature learning. Currently, deep learning models are
 primarily divided into graph-based GNN methods \cite{liu2021combining,liu2021smart,wu2021peculiar,liu2021combining} 
 and word embedding-based methods \cite{chen2023smart,li2018vuldeepecker,yue2020sentiment}. In GNN methods,
  the CGE model represents contract semantics using a graph structure, while AMEVulDetector combines domain 
  expert knowledge with graph convolutional networks to optimize feature extraction. As research advances, 
  Liu et al. introduced the Time Message Propagation (TMP) model, which optimizes node features and temporal 
  propagation mechanisms to address the flatness issue in GNNs. Notably, Qian P et al. \cite{qian2023cross} proposed a 
  cross-modal distillation framework that uses graph attention networks to process semantic graphs of source 
  code and bytecode, enhancing generalization through a dual-teacher network. However, GNN methods heavily 
  rely on the quality of graph construction, limiting their wide application across diverse contract datasets, 
  and facing challenges in information fusion and lightweight scaling. In word embedding-based methods, 
  Lli et al. used Bi-LSTM to extract vulnerability semantic features, capturing long dependencies with 
  relatively low computational demands. The CBGRU model combines Bi-LSTM with convolutional neural networks, 
  integrating multiple word embedding methods to improve detection accuracy. However, this fusion approach 
  requires fine-tuning and the integration of different types 
of word embedding features, which increases training complexity and dependence on hyperparameters.
\par In this paper, we build upon the work of Qian et al. to propose an adaptive attention 
fusion smart contract vulnerability detection model based on data-free distillation. 
Unlike other models that rely on complex cross-modal distillation or CBGRU hybrid models, this 
model eliminates the necessity for large-scale labelled data and solely employs the feature 
generation mechanism to acquire teacher model knowledge. This approach significantly streamlines 
data dependency and model complexity, enhancing efficiency 
and scalability in resource-constrained settings.

\section{Our Approach}
\subsection{Data Preprocessing}
\begin{figure}[htbp]
  \centering
  \includegraphics[width=0.8\textwidth]{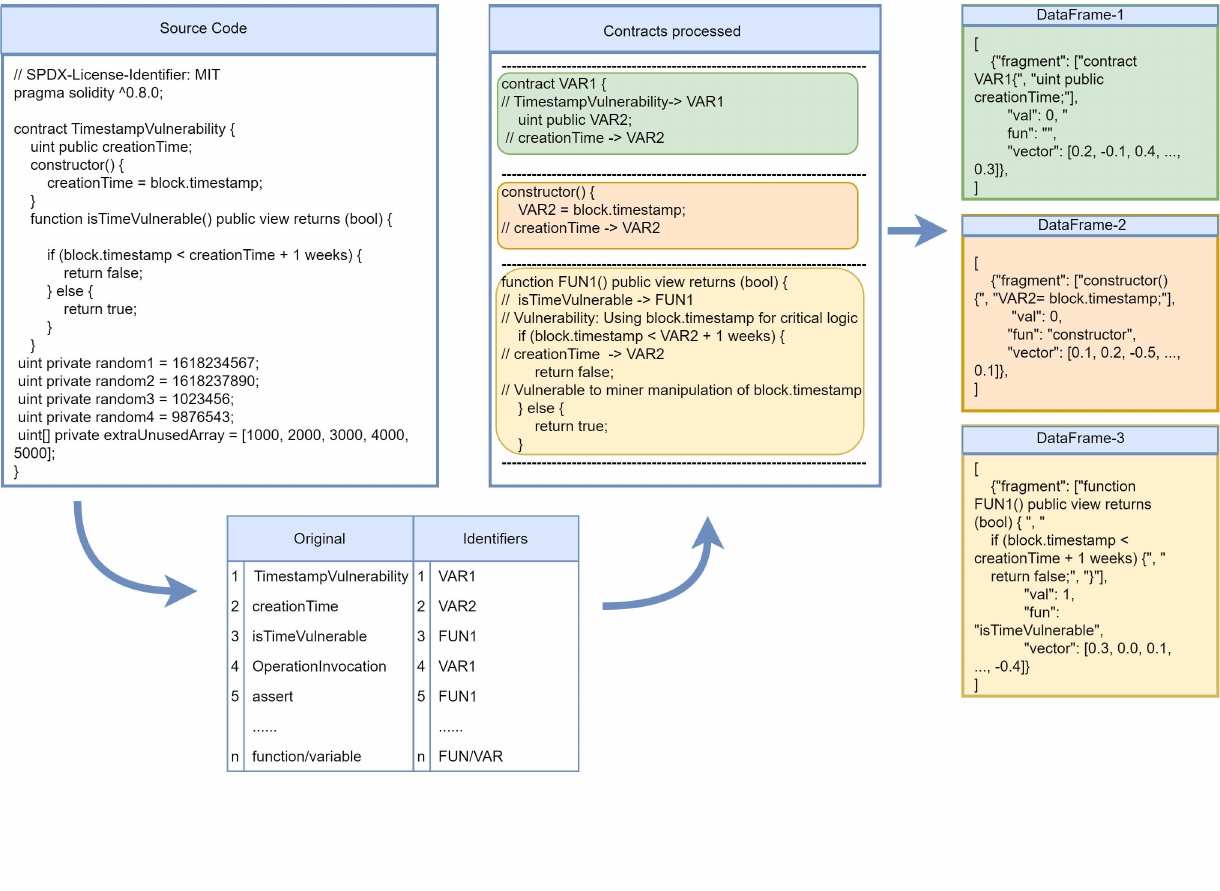}
  \caption{depicts the methodology employed to derive embedding vectors from raw smart contract code through preprocessing and normalization. This process encompasses the following steps: (A) The removal of superfluous information from the source code; (B) The segmentation of pertinent code segments and the annotation of critical vulnerability behaviours; (C) The processing of word embedding and the generation of a matrix.}
  \label{fig1}
\end{figure}

In contrast to conventional textual data, raw smart contracts frequently exhibit intricate 
code structures, encompassing variable declarations, version declarations, function definitions 
for diverse functionalities, and the distinctive coding comment practices of the creators. In 
order to enhance data standardisation and meet the input requirements of neural networks, this 
study employs a preprocessing technique that transforms the main structure of smart contracts 
into a matrix format that is compatible with the model. The data preprocessing workflow is 
depicted in Figure \ref{fig1}. 

\par It is important to acknowledge that out-of-vocabulary words and diverse word forms in 
code have consistently posed a significant challenge for word embedding models. 
To address potential issues in the preprocessing of smart contracts, this study proposes two 
solutions.
\par The CBOW model in Word2Vec \cite{mikolov2013distributed} is used alone: The method captures the corresponding 
semantic information by representing words as fixed-dimensional vectors and calculating 
the distance relationship between the word vectors in space. The experimental results 
demonstrate that Word2Vec is an effective method for capturing the semantic information 
of the code for known vocabulary. Furthermore, it maintains the relative positional 
relationships of phrases and contextual semantics within the contractual structure, thereby 
generating accurate semantic vectors.
\par The CBOW model is combined with the FastText word embedding \cite{bojanowski2017enriching} representation:
 In order to gain further insight into the impact of unlogged words and change-rich word 
 shapes on the generation of vulnerability features, we have combined the CBOW model with 
 FastText for the purpose of vector representation generation. In contrast to Word2Vec, 
 FastText generates word vectors by dividing words into suburbs and combining them. This 
 methodology more accurately captures the internal structure of words, 
rendering it well-suited for addressing out-of-vocabulary words and complex word forms in 
large-scale datasets.
\par
\begin{figure}[htbp]
  \centering
  \includegraphics[width=0.4\textwidth]{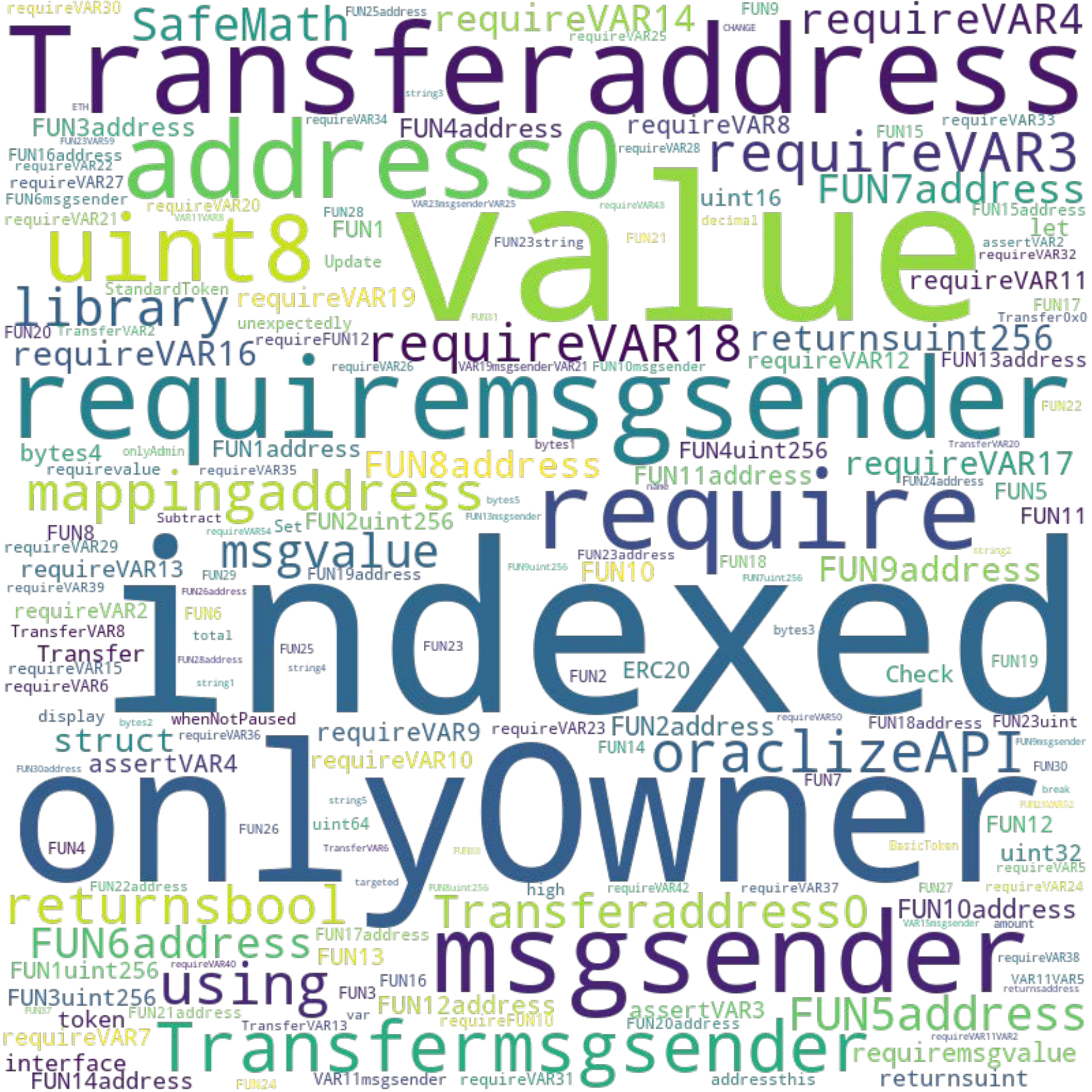}
  \caption{Word cloud map of key semantic distributions in the smart contract vulnerability dataset.}
  \label{fig2}
\end{figure}

 However, the results of our experiments indicated that when the code structure is 
relatively stable, combining the CBOW model with FastText embedding did not result in 
superior outcomes compared to utilising the CBOW model in Word2Vec on its own. It is 
hypothesised that the fixed naming conventions and code syntax structure inherent to the
 process of writing smart contracts result in a lack of significant improvement in the model's 
 understanding of smart contracts when FastText's sapwood decomposition is introduced. 
 Conversely, this may result in the introduction of noise and redundancy in the word embedding 
 analysis, which has a detrimental impact on the model's performance. In order to visually 
 demonstrate the semantic distribution of the smart contract data, we constructed a 
word cloud analysis of key code segments from selected vulnerability datasets, as illustrated 
in Figure \ref{fig2}. 

\par The word cloud illustrates that specific keywords within the contract code, including "transfer"
 (transfer operation), "only Owner" (permission control modifier), "value" (value-related field), 
 and "msg. sender" (caller address), are recurrent. These terms reflect the fundamental 
logic of the contract code and provide an effective data structure for the CBOW word embedding model. 
\par To ensure that semantically similar code words are closer in the vector space and to better 
integrate relative positional information, each word is embedded into a 300-dimensional vector space, 
with positional encoding added as a supplement to the sequence information\cite{vaswani2017attention}. It is represented as $Repeat(X_{input},repeat=K)\in R^{B*(N*K)*C}$, 
where N denotes the sequence length, C denotes the embedding dimension, and K denotes the expansion factor. 
This method helps the model better capture long-range dependencies by expanding the input sequence. 
Additionally, to reduce noise introduced by comments and invalid characters in the contract code, 
we remove these parts during the contract normalization process, leaving only the core keywords. 
This enhances the word embedding layer's ability to capture vulnerability features. For specific 
vulnerability functions, we annotate the function's contextual code using regular expressions and syntax analysis, 
splitting the code line by line to construct finer input units that provide higher granularity for word embedding.

\subsection{Teacher-Student Models}
\begin{figure}[htbp]
  \centering
  \includegraphics[width=1\textwidth]{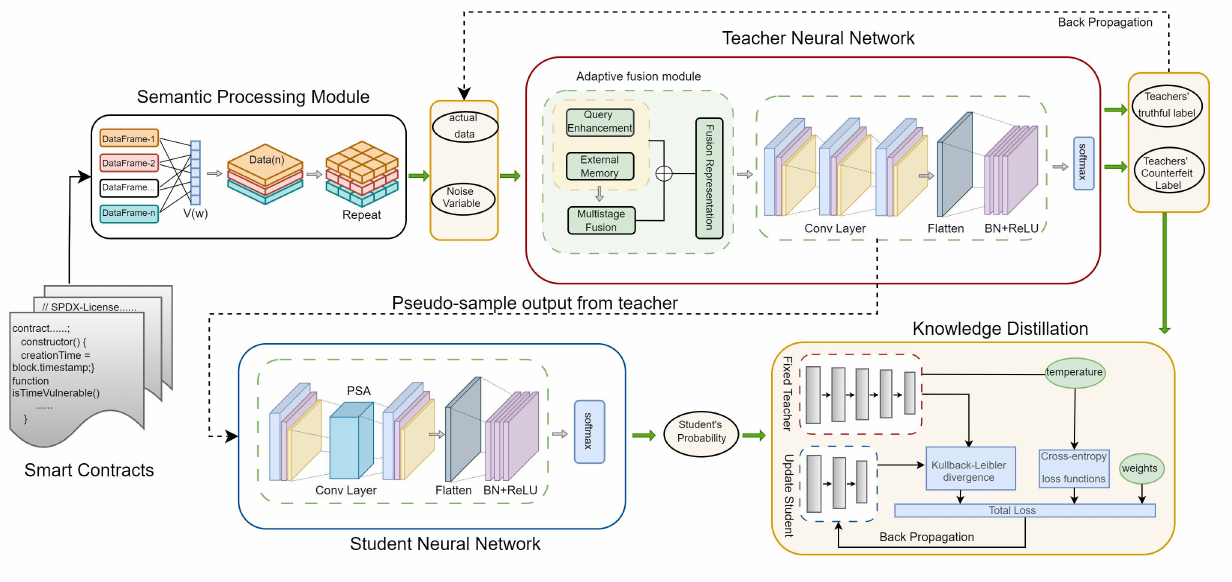}
  \caption{The smart contract vulnerability detection framework in this paper. (A) Semantic Processing Module: Converts the smart contract source code into embedded vector representations and uses positional encoding to enhance vulnerability feature representation. (B) Teacher-Student Neural Network Framework: Consists of a complex teacher network (including the adaptive fusion module) and a student Network.}
  \label{fig3}
\end{figure}

In the data preprocessing phase, the Word2Vec embedding method was employed to extract feature vectors for 
the smart contracts. Subsequently,
 a teacher-student neural network model based on data-free distillation was constructed, as shown in 
 Figure \ref{fig3}.

\par Teacher Model. The teacher model accepts embedded vectors as input and extracts deep features through an 
adaptive attention fusion module, thereby generating global feature representations. A comprehensive account 
can be found in Section 3.4. Subsequently, a three-layer convolutional network (comprising $1\times 3$ convolution 
kernels with filter sizes of 64, 128, and 256) is employed for feature extraction, followed by batch 
normalization, ReLU activation, and max pooling for dimensionality reduction. During the training phase, 
the feature weights extracted from this layer are retained as teacher knowledge and transferred to the student 
model. Ultimately, 
the model generates the teacher's classification results through a fully connected layer with softmax activation.

\par Student Models. The student model employs a lightweight architectural approach, which is a simplified 
version of the teacher model. The student model employs a minimalistic approach by utilising only two 
convolution layers (with filters of 64 and 128, and kernel size of $1\times 3$) for feature extraction. In order 
to enhance feature extraction capabilities and prevent over-simplification, the adaptive attention fusion 
module has been removed, and the PSA attention module \cite{zhang2022epsanet} has been inserted between the convolution layers. 
Subsequently, the feature vectors undergo 
processing through two fully connected layers, and batch normalisation is applied to prevent overfitting.

\subsection{Data-free knowledge distillation}
To ensure that the student model effectively acquires knowledge from the teacher model, we employ a knowledge 
distillation strategy based on prior knowledge for collaborative training, thereby enhancing the training 
performance of the student model. The majority of traditional knowledge distillation methods rely on the 
utilisation of true labels for the purpose of distillation. However, in a data-free distillation scenario, 
the conventional cross-entropy loss function with true labels is not employed as the target. Instead, the 
feature outputs derived from a specific layer of the teacher model are employed as prior knowledge, thereby 
guiding the student in its learning process.

To enhance the diversity and representativeness of training samples, random noise perturbation is introduced, where the random noise variable 
serves as the input for generating pseudo-samples. The noise variable is initialized as $\mathrm{z}{\sim}\mathrm{N}(\mu,\sigma^{2})$, where $\mathbb{Z}\in\mathbb{R}^{\mathrm{d}}$, d denotes the dimensionality of the 
noise variance, and ${\sigma}^{\mathrm{2}}$ represents the variance of the noise, controlling the range of the sampling distribution. Specifically, to ensure that the 
generated pseudo-samples effectively capture the feature distribution of the teacher model, a noise optimization objective function is introduced. 
The Mean Squared Error (MSE) is employed to optimize the noise variance z, driving the activation features $\alpha^{i}_{False}$ extracted from the teacher model from 
pseudo-samples to closely approximate the activation features $\alpha^{i}_{True}$ generated by real data. This optimization is achieved through backpropagation, refining z iteratively.
The optimization process is formalized in Equations (1-3), where $g_{t}$ denotes the activation function output, encapsulating prior knowledge, $f_{t}$ represents 
the mapping function of the teacher model from input to output, $\theta_{t}$ refers to the teacher model parameters, $\eta$ specifies the learning rate controlling the 
update step size, and $\frac{\alpha_{L_{MSE}}}{\alpha_{z}}$ is the gradient of the noise variance. Notably, z is initialized as a noise term sampled from a Gaussian distribution.
\begin{align}
  \mathrm{X_{False}=f_{t}(z;\theta_{t}),\alpha_{False}=g_{t}(X_{False};\theta_{t})} \tag{1} \\
  \mathrm{L_{MSE}=\frac{1}{C}\sum_{i=1}^{c}||\alpha^{i}_{False}-\alpha^{i}_{True}||^{2}}\tag{2} \\
  \mathrm{z\leftarrow z-\eta\frac{\alpha_{L_{MSE}}}{\alpha_z}}\tag{3}
\end{align}
The distillation loss and classification loss are calculated using Kullback-Leibler (KL) divergence and cross-entropy loss functions (CLF), respectively. 
To determine the difference between the pseudo-label $\hat{\mathrm{P}}_{\mathrm{teacher}}$ of the teacher model and the output $P_{student}$ of the student model on the optimized pseudo-sample. This enables 
the student model to emulate the decision-making process of the teacher during training, thereby capturing the intricate knowledge embedded within the teacher 
network. By employing a data-free distillation strategy, the student model can enhance its accuracy and robustness without the necessity of real data labels. 
Ultimately, by weighting and combining the distillation loss and classification loss, the total loss of the student network is obtained, thereby balancing 
the contributions of both to the model training. These are represented by formulas (4-6). In this context, ${P}_{teacher}$ represents the output of the teacher model for the 
i-th feature position, $\hat{P}_{teacher}$ is the pseudo-label 
generated by the teacher model, C denotes the feature dimension size, and $\alpha$ is the weight coefficient that balances the distillation loss and classification loss.

\begin{align}
  \mathrm{\mathrm{L_{KL}(P_{teacher},P_{student})}=\sum_{i=1}^{n}P_{teacher}\log(\frac{\hat{P}_{teacher}}{P_{student}}),P_{student}=\frac{e^{T(i)}}{\sum_{j=1}^{C}e^{T(j)}}}\tag{4} \\
  \mathrm{\mathrm{L_{CLF}(\hat{P}_{teacher},P_{student})}=-\frac{1}{C}\sum_{i=1}^{C}\hat{P}_{teacher}\mathrm{logP_{student}}^{i}}\tag{5} \\
  \mathrm{\mathrm{L_{concat}=\partial L_{KL}(P_{teacher},P_{student})}+(1-\partial)\mathrm{L_{CLF}(\hat{P}_{teacher},P_{student})}}\tag{6}
\end{align}

\subsection{Adaptive Fusion Module}
\begin{figure}[htbp]
 \centering
 \includegraphics[width=1\textwidth]{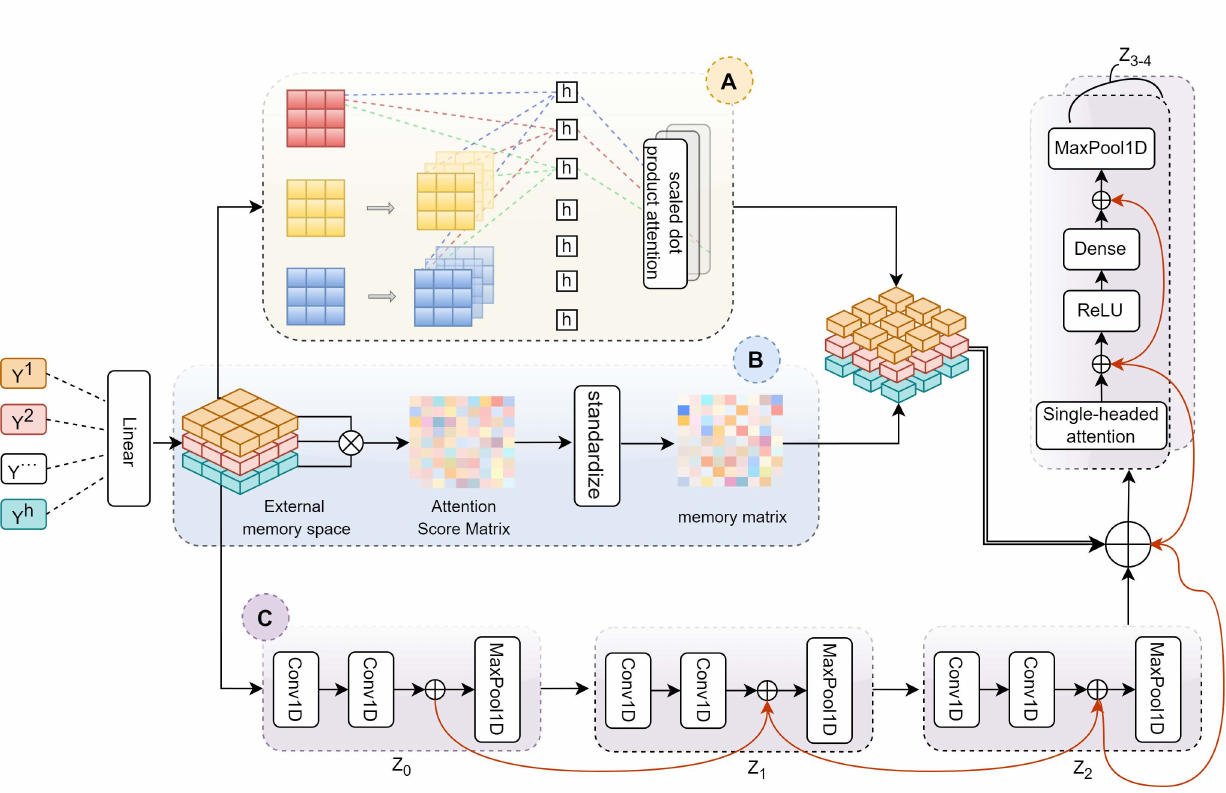}
 \caption{General architecture of the adaptive fusion module in this paper. (A) Multi-dimensional query enhancement mechanism. (B) External memory modelling. (C) Multi-level convolutional fusion enhancement. In the figure $\oplus$ denotes the interaction between the modules and the enhancement operation.}
 \label{fig4}
\end{figure}

The adaptive fusion module, as proposed in this paper, is an attentional mechanism that combines local feature extraction 
with global relationship modelling. Its primary function is to enhance sequence modelling capabilities through 
multi-dimensional query enhancement, external memory modelling, and multi-stage convolutional fusion mechanisms, thereby 
enabling bottom-up, staged feature interaction extraction. During the feature extraction process, the adaptive fusion module 
first proposes using a multi-dimensional query enhancement mechanism to divide the input features into groups. By introducing 
scaled dot-product attention, it represents the queries and keys within each group, enabling fine-grained relationship 
modeling between different channels. Subsequently, the external memory modelling mechanism deploys the augmented query 
capabilities to establish global dependencies across the contract execution stages (i.e. disparate time points in the sequence).
 By employing an external memory space representation matrix to supplement the input sequence features, 
 the model's attention modelling capability for key logical chains is enhanced. Building upon this, the multi-stage 
 convolutional fusion enhancement mechanism combines multi-dimensional query enhancement and external memory modeling. 
 By stacking convolutional layers and pooling operations, it gradually extracts local features and performs multi-scale 
 feature fusion. This effectively improves the model's ability to capture both local features and long-range dependencies, 
 further strengthening its capacity to 
understand and represent complex feature data. The module structure diagram is shown in Figure \ref{fig4}.

The following section provides a comprehensive description of each mechanism.
\par (1)Multi-dimensional query enhancement mechanism
\par Typically, standard query mechanisms process the query, keys, and values of input features 
  in a uniform manner, generating weight matrices and weighting and summing them through similarity
   calculations. The Scaled Dot-Product Attention \cite{vaswani2017attention} is an illustrative example. The similarity 
   matrix $QK^{T}$ is computed by the dot product of the query Q and key K. This matrix is scaled by a
    factor $\sqrt{d_k}$ and passed through a softmax function to obtain the attention 
    weights for the keys, 
    where $d_{k}$ is the key dimension. The weighted sum of the values is then calculated as $\frac{QK^T}{\sqrt{d_k}}*V$, where V 
    is the value matrix. However, the standard query mechanism typically involves single-level 
    feature interaction, which renders it less suitable for downstream vector tasks. As previously 
    stated, we have developed a feature enhancement module that integrates feature grouping 
    modelling with attention distribution calculation. This is known as the Multi-Dimensional 
    Query Enhancement Mechanism. The underlying computational process is illustrated in equations 
    (7-9). In the above equation, $\mathrm{Q^h,K^h,V^h,Z^h\in R^{B*(N*K)*C*numhead},X^{^{\prime}}\in R^{B*(N*K)*C},W^0\in R^{c*c}}$, numhead denotes the number 
    of heads, $Z^h$ denotes the output 
    feature matrix of each head, and $\mathrm{X}^{\prime}$ denotes the new aggregation result. Firstly, the input 
    feature matrix is mapped to obtain the query $\mathrm{Q^{(h)}}$, key $\mathrm{K}^{(\mathrm{h})}$, 
    and value $\mathrm{V}^{(\mathrm{h})}$ representations along the 
    channel dimension. The $\mathrm{K}^{(\mathrm{h})}$ and $\mathrm{V}^{(\mathrm{h})}$ are grouped based on a specified number of groups. Within each 
    group, an independent group expansion mechanism is applied to map the features back to the 
    multi-head dimensions, ensuring that different attention heads share the fine-grained feature 
    distribution within the group. Secondly, scaled dot-product attention scores are computed for 
    each group, generating an attention weight matrix based on the matching distribution between 
    queries and keys within the group. This matrix is then normalized and combined to obtain the 
    weighted sum, as shown in equation (5), enabling the adaptive fusion of global information and 
    local relationships. Finally, the output features of each 
  attention head are projected through a projection matrix $W^{0}$ to form a new aggregated 
  representation $\mathrm{X}^{\prime}$.
  \begin{align}
    \mathrm{Q^{(h)}=split(Q),K^{(h)}=splitnum(K),V^{(h)}=splitnum(V)}\tag{7} \\
    \mathrm{Z^h}=\mathrm{ATT}(\mathrm{Q}^{(h)},\mathrm{K}^{(h)},\mathrm{V}^{(h)})=\mathrm{softmax}(\frac{\mathrm{Q}^{(h)}*(\mathrm{K}^{(h)})^\mathrm{T}}{\sqrt{\mathrm{d_h}}})*\mathrm{V}^{(h)}\tag{8} \\
    \mathrm{X^{^{\prime}}=Concat(Z^{(1)},~Z^{(2)},~\cdots,~Z^{(h)})*W^{0}}\tag{9}
  \end{align}

  \par In comparison to conventional query mechanisms, this module prioritises the capture of local 
  features, thereby offering more robust channel feature representation capabilities. 
  This results in a notable enhancement in the efficiency with which features can be utilised 
  for downstream tasks.

  \par (2)External memory modelling
  \par Building upon the multi-dimensional query enhancement mechanism, we 
  employ external memory modelling to provide supplementary global features for 
  the current input sequence. In particular, the external memory modelling mechanism enhances the 
  model's ability to model long-distance sequences, thereby storing global information across 
  contract execution steps. This allows each output to be combined with previously computed 
  aggregated representation features to further construct global information about key operational 
  steps and feature dimensions. In each layer, the input feature $\mathrm{Y^{h}}$ is mapped through the dense layer 
  into an external memory representation space $\mathrm{M}^{B*(N*K)*S}$, where S is the dimension of the memory space. 
  Then, the attention scores of the augmented features of the current layer are computed with the 
  external memory matrix to obtain the attention score matrix A, as shown 
  in equation (10). Where $Y_i^h$ is the i-th element of the input feature $Y^{h}$ and $\mathrm{M_i}$ is the i-th element in 
  the external memory matrix M.
  \begin{align}
    \mathrm{A_i=\frac{\exp(Y_i^h*M_i)}{\sum_i\exp(Y_i^h*M_i)}}\tag{10}
  \end{align}
  \par In the pooling layer, the retrieved memories are constrained to be weighted by the obtained
   attention score A to generate the updated memory matrix. And the enhanced features are projected back 
  to the original feature space using the back-mapping mechanism of the memory space to obtain $\mathrm{Y^{'}}$.
   As shown in equation (11).
   \begin{align}
    M_{new}=update(\mathrm{Y^h},M)=\sum_i\mathrm{A_i}*\mathrm{M_i}\tag{11}
  \end{align}
\par The enhanced memory features facilitate the aggregation of global information, thereby 
reducing the likelihood of over-localisation or an excessive focus on local details. This is
 achieved by effectively capturing long-range dependencies, which can be challenging to achieve
  when relying solely on the multi-dimensional query enhancement mechanism.

\par (3)Multi-stage convolutional fusion enhancement mechanism
\par In complex data processing, compared to neural network structures that tend to overfit the 
training samples, our proposed multi-stage convolutional fusion enhancement mechanism demonstrates 
superior ability in reinforcing relational modeling. This approach is theoretically grounded. 
To illustrate, the CoAtNet network \cite{dai2021coatnet} utilises a pyramid-like architectural configuration centred 
upon convolutional stride downsampling and multi-head attention, which serves to augment both 
local feature extraction and global feature modelling capabilities. The vast parameter space and 
intricate convolutional configuration render the network susceptible to overfitting when confronted 
with multidimensional data, impeding the ability to discern the intrinsic patterns of the data.
 Accordingly, this paper builds upon the CoAtNet design, implements a stage-by-stage processing 
 approach, and incorporates the MaxPool1D downsampling operation within the multi-stage convolutional 
 fusion enhancement module. This operation extracts the local maxima through the pooling mechanism, 
 thereby reducing the dimensionality and better adapting the data to the feature extraction needs of 
 the model. The MaxPool1D downsampled features are input in a compact form 
to the external memory modelling and multidimensional query mechanism, thus reducing the burden 
of subsequent computation. 
\par Specifically, during the convolutional feature extraction phase, local context spatial 
features are extracted by stacking convolutions. The maximum pooling layer is then applied to 
reduce the time step length, 
thereby further enhancing the model's ability to perceive local regions, as illustrated in 
Equation (12). Where $\mathbb{Z}^0\in\mathbb{R}^{\mathrm{repeat}},\mathbb{Z}^{(1)}\in\mathbb{R}^{B*\frac{N*K}{2}*C}$.
\begin{align}
  \mathrm{Z^0=Conv1D~(X)~,~Z^{(1)}~=MBConv(Z^0)~,~Z^{(i+1)}~=MaxPool1D~(MLP(MBConv(Z^{(i)})))}\tag{12}
\end{align}
\par In the feature enhancement and global relationship modeling phase, this paper proposes 
combining single-head attention with external memory modeling and multi-dimensional query 
mechanisms. Through the step-by-step extraction of precise features and convolutions, this 
approach replaces the subspace decomposition ability of multi-head attention in CoAtNet, 
overcoming the limitations of single-head attention. Compared to CoAtNet, this method significantly 
optimizes the number of parameters and achieves a better balance between feature richness and 
computational speed. This method employs the use of $\mathrm{Z}^{(\mathrm{i}+1)}$ as the interaction between the query, key, 
and value. The attention mechanism serves to fuse the output features and subsequently 
map them to the target channel dimensions. The final output features are then concatenated 
with the multi-dimensional enhancement mechanism and external memory modelling features, 
resulting in the generation of higher-level semantic information $\mathrm{F=Concat(X^{^{\prime}},Y^{^{\prime}},Z^{(i+1)})}$. 
This approach facilitates a hierarchical fusion of features from local to global, 
as illustrated in Equation (13).

\begin{align}
  \mathrm{Z^{(i+1)}=MaxPool1D~(MLP(Atten(Z^{(i)})))=MaxPool1D~(MLP(softmax(\frac{Q*(K)^T}{\sqrt{d_K}})*V)))}\tag{13}
\end{align}

\section{Experimental}
\subsection{Data set}
In order to guarantee the standardisation of data in the field of smart contract detection, we take the
 current public dataset published in the renowned paper \cite{qian2023cross} in the field of smart contract vulnerability 
 detection as our point of departure. We then merge the current extensive smart contract vulnerability dataset 
 SmartBugs Wild \cite{durieux2020empirical}, which is a recently released Solidity language file, for manual tagging. The dataset was 
 constructed with the objective of encompassing a comprehensive array of smart contract vulnerabilities, 
 encompassing five distinct categories: reentrancy vulnerability, timestamp vulnerability, delegatecall 
 vulnerability, integer overflow/underflow vulnerability, and CDAV (Contract Deployment Address) vulnerability. 
 In light of the disparate distribution of vulnerability rates within the dataset, an undersampling methodology 
 was employed. The number of samples randomly selected from the negative categories was equal to the number 
 of samples randomly selected from the positive categories, thus ensuring a balanced distribution of sample 
 categories. The data preprocessing method proposed in this paper is the primary method used for labelling 
 and processing the dataset. The specific steps are detailed in Section 3.1. Additionally, the results presented 
 in this paper are primarily compared with the performance metrics published in the original paper that provides 
 the dataset on GitHub, which were then experimented with as supplementary results.
 Consequently, the first four vulnerabilities have been selected as the performance metrics of this model for
  validation.

  \subsection{Experimental setup}
  Experimental environment. The experimental environment is described below. All experiments were conducted on 
  a computer system comprising an Intel Xeon Gold 5115 central processing unit, a NVIDIA Tesla T4 graphical 
  processing unit and 16 gigabytes of random-access memory. The model network proposed in this paper is implemented 
  using 
  the Keras and TensorFlow frameworks, with the code implementation based on Python 3.7 and the operating system
   Ubuntu.
\par Experimental parameters. In order to train our model, we determined the learning rate settings by means of 
a grid search, with the learning rate adjusted between 1e-4 and 1e-2. During the training phase, the optimiser 
was selected to be the Adam optimiser, with the AMSGrad variant enabled, in order to enhance the stability of 
the training process. During the distillation phase, the optimiser was configured to utilise stochastic gradient 
descent (SGD) and a pre-learning rate scheduler, with the batch size varying between 8, 16, 32, 64, and 128. 
Furthermore, as the student model enhances its detection capabilities through the transfer of knowledge, we 
designated the weighting parameter $\alpha$ as a hyperparameter, with $\alpha$ selected within the range of 0 to 1. Following 
a series of experiments and parameter tuning, the optimal default parameter settings were identified as a learning 
rate of 1e-3, a batch size of 64, an initial vector dimensionality of 300, SGDs with an Adam and momentum parameter
 of 0.9 for the optimisers, and a weight parameter of 0.2 for the $\alpha$. For each dataset, 80\% were randomly selected as the
 training set and 20\% as the test set. Each set of experiments was repeated five times to obtain the average results.

\par Evaluation of indicators. Four evaluation criteria have been selected to assess the performance of the model in 
the experiment. These are accuracy, recall, precision and F1 value, as illustrated in Equations 14-17.
\begin{align}
  \mathrm{Accuracy}=\frac{\mathrm{TP+TN}}{\mathrm{TP+TN+FP+FN}} \tag{14} \\
  \mathrm{Precision}=\frac{\mathrm{TP}}{\mathrm{TP}+\mathrm{FP}} \tag{15} \\
  \mathrm{Recall}=\frac{\mathrm{TP}}{\mathrm{TP}+\mathrm{FN}} \tag{16} \\
  \mathrm{F1}=2\times\frac{\mathrm{Precision}\times\mathrm{Recall}}{\mathrm{Precision}+\mathrm{Recall}} \tag{17}
\end{align}
\par In this context, TP represents the number of contracts that were correctly identified as vulnerable, TN denotes 
the number of contracts that were correctly identified as not vulnerable, FP signifies the number of contracts that 
were incorrectly 
identified as vulnerable, and FN represents the number of contracts that were incorrectly identified as not 
vulnerable.

\subsection{Experimental results and analysis}
\subsubsection{Performance Comparisons}
In order to evaluate the efficacy of the model proposed in this paper in detecting the four vulnerabilities,
 a number of traditional vulnerability detection tools were selected for comparison, including sFuzz \cite{nguyen2020sfuzz}, Oyente, 
 Mythril, Osiris \cite{torres2018osiris}, SmartCheck \cite{tikhomirov2018smartcheck}, and Slither \cite{feist2019slither}. 
The vulnerability dataset was subjected to a benchmarking process, the results of which are presented in Table 1.

\begin{table}[!ht] 
  \centering
  \label{tab:1}
  \caption{Performance comparison of existing smart contract vulnerability detection tools. It contains 6 tools. None indicates that the corresponding tool does not support detection of this vulnerability type.}
  \arrayrulecolor{black}
  \resizebox{\textwidth}{!}{
  \begin{tabular}{|l||llll||llll||llll||llll|}
    \hline
    \multirow{2}{*}{\begin{tabular}[c]{@{}l@{}}Static\\ Based\end{tabular}} & \multicolumn{4}{l||}{Reentrancy} & \multicolumn{4}{l||}{Timestamp} & \multicolumn{4}{l||}{Delegatecall} & \multicolumn{4}{l||}{IntegerOverflow/Underflow} \\ \cline{2-17} 
     & \multicolumn{1}{l|}{ACC} & \multicolumn{1}{l|}{Recall} & \multicolumn{1}{l|}{Precision} & F1 Score & \multicolumn{1}{l|}{ACC} & \multicolumn{1}{l|}{Recall} & \multicolumn{1}{l|}{Precision} & F1 Score & \multicolumn{1}{l|}{ACC} & \multicolumn{1}{l|}{Recall} & \multicolumn{1}{l|}{Precision} & F1 Score & \multicolumn{1}{l|}{ACC} & \multicolumn{1}{l|}{Recall} & \multicolumn{1}{l|}{Precision} & F1 Score \\ \hline
    SFuzz\cite{nguyen2020sfuzz} & \multicolumn{1}{l|}{55.69} & \multicolumn{1}{l|}{14.95} & \multicolumn{1}{l|}{10.88} & 12.59 & \multicolumn{1}{l|}{33.41} & \multicolumn{1}{l|}{27.01} & \multicolumn{1}{l|}{23.15} & 24.93 & \multicolumn{1}{l|}{64.37} & \multicolumn{1}{l|}{47.22} & \multicolumn{1}{l|}{58.62} & 52.31 & \multicolumn{1}{l|}{45.50} & \multicolumn{1}{l|}{25.97} & \multicolumn{1}{l|}{25.88} & 25.92 \\ \hline
    Oyente\cite{luu2016making} & \multicolumn{1}{l|}{65.07} & \multicolumn{1}{l|}{63.02} & \multicolumn{1}{l|}{46.56} & 53.55 & \multicolumn{1}{l|}{68.29} & \multicolumn{1}{l|}{57.97} & \multicolumn{1}{l|}{61.04} & 59.47 & \multicolumn{1}{l|}{None} & \multicolumn{1}{l|}{None} & \multicolumn{1}{l|}{None} & None & \multicolumn{1}{l|}{69.71} & \multicolumn{1}{l|}{57.55} & \multicolumn{1}{l|}{58.05} & 57.80 \\ \hline
    Mythril\cite{ruskin1980mythril} & \multicolumn{1}{l|}{64.27} & \multicolumn{1}{l|}{75.51} & \multicolumn{1}{l|}{42.86} & 54.68 & \multicolumn{1}{l|}{62.40} & \multicolumn{1}{l|}{49.80} & \multicolumn{1}{l|}{57.50} & 53.37 & \multicolumn{1}{l|}{75.06} & \multicolumn{1}{l|}{62.07} & \multicolumn{1}{l|}{72.30} & 66.80 & \multicolumn{1}{l|}{None} & \multicolumn{1}{l|}{None} & \multicolumn{1}{l|}{None} & None \\ \hline
    Osiris\cite{torres2018osiris} & \multicolumn{1}{l|}{56.73} & \multicolumn{1}{l|}{63.88} & \multicolumn{1}{l|}{40.94} & 49.90 & \multicolumn{1}{l|}{66.83} & \multicolumn{1}{l|}{55.42} & \multicolumn{1}{l|}{59.26} & 57.28 & \multicolumn{1}{l|}{None} & \multicolumn{1}{l|}{None} & \multicolumn{1}{l|}{None} & None & \multicolumn{1}{l|}{68.41} & \multicolumn{1}{l|}{34.18} & \multicolumn{1}{l|}{60.83} & 43.77 \\ \hline
    SmartCheck\cite{tikhomirov2018smartcheck} & \multicolumn{1}{l|}{54.65} & \multicolumn{1}{l|}{16.34} & \multicolumn{1}{l|}{45.71} & 24.07 & \multicolumn{1}{l|}{47.73} & \multicolumn{1}{l|}{79.34} & \multicolumn{1}{l|}{47.89} & 59.73 & \multicolumn{1}{l|}{62.41} & \multicolumn{1}{l|}{56.21} & \multicolumn{1}{l|}{45.56} & 50.33 & \multicolumn{1}{l|}{53.91} & \multicolumn{1}{l|}{68.54} & \multicolumn{1}{l|}{42.81} & 52.70 \\ \hline
    Slither\cite{feist2019slither} & \multicolumn{1}{l|}{74.02} & \multicolumn{1}{l|}{73.50} & \multicolumn{1}{l|}{74.44} & 73.97 & \multicolumn{1}{l|}{68.52} & \multicolumn{1}{l|}{67.17} & \multicolumn{1}{l|}{69.27} & 68.20 & \multicolumn{1}{l|}{68.97} & \multicolumn{1}{l|}{52.27} & \multicolumn{1}{l|}{70.12} & 59.89 & \multicolumn{1}{l|}{None} & \multicolumn{1}{l|}{None} & \multicolumn{1}{l|}{None} & None \\ \hline
  \end{tabular}
  }
  \arrayrulecolor{black}
  \end{table}

\par The findings indicate that the existing conventional vulnerability detection tools, including SFuzz, 
Oyente, Mythril and other similar tools, are unable to achieve a high level of detection for all indicators,
 particularly in the case of integer overflow and Delegatecall vulnerabilities. This deficiency is more 
 pronounced in these areas. It is postulated that this phenomenon is primarily attributable to the dearth of 
 modelling awareness pertaining to the dynamic execution environment among these tools. In contrast, 
 sophisticated vulnerabilities frequently necessitate the analysis of numerous interactions within the 
 contract, a task which the aforementioned tools are unable to accomplish. For example, SFuzz achieves a 
 recall rate of only 25.97\% in detecting integer overflow vulnerabilities, indicating a high false positive 
 rate and significant misclassification of vulnerabilities. Compared to traditional vulnerability detection 
 tools, our model improves detection metrics by an average of 15\%-30\% across
 four vulnerability datasets compared to the best existing detection tools, significantly outperforming
  current solutions.

\par We also compare the performance with eight deep learning model-based detection methods such as
 Vanilla-RNN \cite{tann2018towards}, ReChecker \cite{qian2020towards}, GCN \cite{kipf2016semi}, and TMP. The results are shown in Table 2, with the best 
 results in bold. As can be seen from the table, the proposed model in this paper outperforms the baseline 
 of the current best deep learning based detection methods in terms of metrics. The average performance on 
 each dataset is 
improved by 5.73\% (Reentrancy), 0.54\% (Timestamp), 7.93\% (Delegatecall), and 5.66\% (IntegerOverflow/Underflow).

\begin{table}[!ht] 
  \centering
  \label{tab:2}
  \caption{Comparison of the performance of this paper's model with eight current mainstream deep learning-based smart contract detection models. The detection metrics include Accuracy (ACC), Recall, Precision, and F1 Score.}
  \arrayrulecolor{black}
  \resizebox{\textwidth}{!}{
  \begin{tabular}{|l||llll||llll||llll||llll|}
    \hline
    \multirow{2}{*}{\begin{tabular}[c]{@{}l@{}}Deep\\ Learn\end{tabular}} & \multicolumn{4}{l||}{Reentrancy} & \multicolumn{4}{l||}{Timestamp} & \multicolumn{4}{l||}{Delegatecall} & \multicolumn{4}{l||}{IntegerOverflow/Underflow} \\ \cline{2-17} 
     & \multicolumn{1}{l|}{ACC} & \multicolumn{1}{l|}{Recall} & \multicolumn{1}{l|}{Precision} & F1 Score & \multicolumn{1}{l|}{ACC} & \multicolumn{1}{l|}{Recall} & \multicolumn{1}{l|}{Precision} & F1 Score & \multicolumn{1}{l|}{ACC} & \multicolumn{1}{l|}{Recall} & \multicolumn{1}{l|}{Precision} & F1 Score & \multicolumn{1}{l|}{ACC} & \multicolumn{1}{l|}{Recall} & \multicolumn{1}{l|}{Precision} & F1 Score \\ \hline
    Vanilla-RNN\cite{tann2018towards} & \multicolumn{1}{l|}{65.09} & \multicolumn{1}{l|}{72.89} & \multicolumn{1}{l|}{67.39} & 70.03 & \multicolumn{1}{l|}{64.41} & \multicolumn{1}{l|}{65.17} & \multicolumn{1}{l|}{64.16} & 64.66 & \multicolumn{1}{l|}{64.33} & \multicolumn{1}{l|}{67.26} & \multicolumn{1}{l|}{63.77} & 65.47 & \multicolumn{1}{l|}{68.12} & \multicolumn{1}{l|}{70.19} & \multicolumn{1}{l|}{67.00} & 68.56 \\ \hline
    ReChecker\cite{qian2020towards} & \multicolumn{1}{l|}{70.95} & \multicolumn{1}{l|}{72.92} & \multicolumn{1}{l|}{70.15} & 71.51 & \multicolumn{1}{l|}{66.65} & \multicolumn{1}{l|}{54.53} & \multicolumn{1}{l|}{73.37} & 62.56 & \multicolumn{1}{l|}{67.98} & \multicolumn{1}{l|}{70.66} & \multicolumn{1}{l|}{66.47} & 68.50 & \multicolumn{1}{l|}{70.49} & \multicolumn{1}{l|}{71.59} & \multicolumn{1}{l|}{70.56} & 71.07 \\ \hline
    GCN\cite{kipf2016semi} & \multicolumn{1}{l|}{73.21} & \multicolumn{1}{l|}{73.18} & \multicolumn{1}{l|}{74.47} & 73.82 & \multicolumn{1}{l|}{75.91} & \multicolumn{1}{l|}{77.55} & \multicolumn{1}{l|}{74.93} & 76.22 & \multicolumn{1}{l|}{65.76} & \multicolumn{1}{l|}{69.74} & \multicolumn{1}{l|}{69.01} & 69.37 & \multicolumn{1}{l|}{67.53} & \multicolumn{1}{l|}{70.93} & \multicolumn{1}{l|}{69.52} & 70.22 \\ \hline
    TMP\cite{liu2021combining} & \multicolumn{1}{l|}{76.45} & \multicolumn{1}{l|}{75.30} & \multicolumn{1}{l|}{76.04} & 75.67 & \multicolumn{1}{l|}{78.84} & \multicolumn{1}{l|}{76.09} & \multicolumn{1}{l|}{78.68} & 77.36 & \multicolumn{1}{l|}{69.11} & \multicolumn{1}{l|}{70.37} & \multicolumn{1}{l|}{68.18} & 69.26 & \multicolumn{1}{l|}{70.85} & \multicolumn{1}{l|}{69.47} & \multicolumn{1}{l|}{70.26} & 69.86 \\ \hline
    AME\cite{liu2021smart} & \multicolumn{1}{l|}{81.06} & \multicolumn{1}{l|}{78.45} & \multicolumn{1}{l|}{79.62} & 79.03 & \multicolumn{1}{l|}{82.25} & \multicolumn{1}{l|}{80.26} & \multicolumn{1}{l|}{81.42} & 80.84 & \multicolumn{1}{l|}{72.85} & \multicolumn{1}{l|}{69.40} & \multicolumn{1}{l|}{70.25} & 69.82 & \multicolumn{1}{l|}{73.24} & \multicolumn{1}{l|}{71.59} & \multicolumn{1}{l|}{71.36} & 71.47 \\ \hline
    CBGRU\cite{zhang2022cbgru} & \multicolumn{1}{l|}{80.49} & \multicolumn{1}{l|}{83.33} & \multicolumn{1}{l|}{84.04} & 83.68 & \multicolumn{1}{l|}{73.22} & \multicolumn{1}{l|}{78.96} & \multicolumn{1}{l|}{71.11} & 74.83 & \multicolumn{1}{l|}{87.80} & \multicolumn{1}{l|}{81.70} & \multicolumn{1}{l|}{90.54} & 85.89 & \multicolumn{1}{l|}{79.89} & \multicolumn{1}{l|}{\textbf{90.68}} & \multicolumn{1}{l|}{77.37} & 83.49 \\ \hline
    SMS\cite{qian2023cross} & \multicolumn{1}{l|}{83.85} & \multicolumn{1}{l|}{77.48} & \multicolumn{1}{l|}{79.46} & 78.46 & \multicolumn{1}{l|}{89.77} & \multicolumn{1}{l|}{91.09} & \multicolumn{1}{l|}{89.15} & 90.11 & \multicolumn{1}{l|}{78.82} & \multicolumn{1}{l|}{73.69} & \multicolumn{1}{l|}{76.97} & 75.29 & \multicolumn{1}{l|}{79.36} & \multicolumn{1}{l|}{72.98} & \multicolumn{1}{l|}{78.14} & 75.47 \\ \hline
    DMT\cite{qian2023cross} & \multicolumn{1}{l|}{89.42} & \multicolumn{1}{l|}{81.06} & \multicolumn{1}{l|}{83.62} & 82.32 & \multicolumn{1}{l|}{94.58} & \multicolumn{1}{l|}{\textbf{96.39}} & \multicolumn{1}{l|}{93.60} & 94.97 & \multicolumn{1}{l|}{85.64} & \multicolumn{1}{l|}{74.32} & \multicolumn{1}{l|}{85.44} & 79.49 & \multicolumn{1}{l|}{82.76} & \multicolumn{1}{l|}{77.93} & \multicolumn{1}{l|}{84.61} & 81.13 \\ \hline
    STip(Pre) & \multicolumn{1}{l|}{85.41} & \multicolumn{1}{l|}{83.33} & \multicolumn{1}{l|}{84.43} & 83.87 & \multicolumn{1}{l|}{93.58} & \multicolumn{1}{l|}{91.98} & \multicolumn{1}{l|}{94.39} & 93.17 & \multicolumn{1}{l|}{90.24} & \multicolumn{1}{l|}{91.35} & \multicolumn{1}{l|}{90.34} & 90.84 & \multicolumn{1}{l|}{79.55} & \multicolumn{1}{l|}{79.29} & \multicolumn{1}{l|}{81.46} & 80.36 \\ \hline
    STip(Post) & \multicolumn{1}{l|}{\textbf{89.58}} & \multicolumn{1}{l|}{\textbf{89.58}} & \multicolumn{1}{l|}{\textbf{89.74}} & \textbf{89.65} & \multicolumn{1}{l|}{\textbf{95.12}} & \multicolumn{1}{l|}{94.92} & \multicolumn{1}{l|}{\textbf{96.15}} & \textbf{95.55} & \multicolumn{1}{l|}{\textbf{95.73}} & \multicolumn{1}{l|}{\textbf{92.07}} & \multicolumn{1}{l|}{\textbf{95.73}} & \textbf{93.86} & \multicolumn{1}{l|}{\textbf{85.55}} & \multicolumn{1}{l|}{85.03} & \multicolumn{1}{l|}{\textbf{86.11}} & \textbf{85.56} \\ \hline
  \end{tabular}
  }
  \arrayrulecolor{black}
  \end{table}

\par In relatively simple datasets such as reentrancy and timestamp vulnerabilities, RNN-based neural network models 
tend to suffer from the vanishing gradient problem when dealing with long sequence data, resulting in generally lower 
detection efficiency compared to other models. In contrast to the RNN and ReChecker models, our model achieved accuracy 
rates of 89.74\% and 82.45\%, respectively, with F1 scores improving by an average of 20\%-30\%. The graph neural 
network-based TMP model lacks the extraction of fine-grained effective features, although it adds a temporal message 
propagation network to capture global feature information. This results in the model's performance being lower than 
STip's metrics of 13.3\% and 16.28\% for accuracy on the reentrant vulnerability dataset and timestamp vulnerability
dataset, respectively. In addition, it is our contention that the model put forth in this paper, which incorporates 
knowledge distillation and an adaptive fusion module, is capable of effectively extracting pivotal features in the context 
of complex vulnerability detection. This feature enhances the model's ability to recognise complex patterns, particularly 
in vulnerability datasets that require high-precision detection, and offers significant advantages. The experimental 
results demonstrate that the model proposed in this paper, which is based on the data-free knowledge refinement method, 
performs well in detecting delegated call vulnerabilities with an accuracy of 95.73\%. This is approximately 10\% higher 
than the current state-of-the-art DMT model, which is based on the multi-model knowledge refinement approach. This suggests
 that the adaptive fusion module proposed in this paper is more effective in extracting complex feature information from 
 smart contracts, thereby improving the model's accuracy in vulnerability detection tasks. In the detection of integer 
 overflow vulnerabilities, although the STip model shows a 5.65\% decrease compared to the CBGRU model, which has the 
 highest recall, it still outperforms in other metrics, maintaining a strong advantage. Specifically, the STip model 
 demonstrates an accuracy of 85.55\%, a precision of 86.11\%, and an F1 score of 85.56\%. At the same time, after the 
 experiment we found that other models have a large difference too much in the four indicators. For example, the accuracy 
 of the SMS model is 79.36\% while the recall is 72.98\%, a difference of about 7 percentage points. In contrast, our model
  is relatively consistent in the performance 
of the metrics, with small differences in the metrics, demonstrating the strong stability and robustness of the model.

\par To further validate the effectiveness of teacher-student model distillation, an additional experiment was conducted 
in which the student model was trained independently using only the original data, without any distillation knowledge. 
Further details on the experimental results can be found in Section 4.3.3. It is noteworthy that in the absence of 
knowledge regarding the distillation process, the evaluation metrics of the student model exhibited a decline of 4-6 
percentage points on average across the four vulnerabilities. 
This decline was particularly pronounced in the case of integer overflow vulnerabilities, where the decline reached 6\%.

\subsubsection{Ablation Experiment}
In order to ascertain the impact of the multi-dimensional query mechanism, external memory modelling and multi-stage 
convolutional fusion enhancement mechanism on the model within the adaptive fusion module, we have devised ablation 
experiments for each of these three modules. Table \ref{tab:3} illustrates the impact of model distillation on the evaluation metrics, specifically the removal 
of the three aforementioned mechanisms. The notation "w/o" indicates the removal of the corresponding mechanism operation.

\begin{table}[htbp] 
  \centering
  \label{tab:3}
  \caption{Comparison of the performance of distilled STip models in ablation experiments}
  \arrayrulecolor{black}
  \resizebox{\textwidth}{!}{
    \begin{tabular}{|l||llll||llll||llll||llll|}
      \hline
      \multirow{2}{*}{Training Strategy} & \multicolumn{4}{l||}{Reentrancy} & \multicolumn{4}{l||}{Timestamp} & \multicolumn{4}{l||}{Delegatecall} & \multicolumn{4}{l||}{IntegerOverflow/Underflow} \\ \cline{2-17} 
       & \multicolumn{1}{l|}{ACC} & \multicolumn{1}{l|}{Recall} & \multicolumn{1}{l|}{Precision} & F1 Score & \multicolumn{1}{l|}{ACC} & \multicolumn{1}{l|}{Recall} & \multicolumn{1}{l|}{Precision} & F1 Score & \multicolumn{1}{l|}{ACC} & \multicolumn{1}{l|}{Recall} & \multicolumn{1}{l|}{Precision} & F1 Score & \multicolumn{1}{l|}{ACC} & \multicolumn{1}{l|}{Recall} & \multicolumn{1}{l|}{Precision} & F1 Score \\ \hline
      STip(Post) & \multicolumn{1}{l|}{\textbf{89.58}} & \multicolumn{1}{l|}{\textbf{89.58}} & \multicolumn{1}{l|}{\textbf{89.74}} & \textbf{89.65} & \multicolumn{1}{l|}{\textbf{95.12}} & \multicolumn{1}{l|}{\textbf{94.92}} & \multicolumn{1}{l|}{\textbf{96.15}} & \textbf{95.55} & \multicolumn{1}{l|}{\textbf{95.73}} & \multicolumn{1}{l|}{\textbf{92.07}} & \multicolumn{1}{l|}{\textbf{95.73}} & \textbf{93.86} & \multicolumn{1}{l|}{\textbf{85.55}} & \multicolumn{1}{l|}{\textbf{85.03}} & \multicolumn{1}{l|}{\textbf{86.11}} & \textbf{85.56} \\ \hline
      Query Enhancement-w/o & \multicolumn{1}{l|}{89.06} & \multicolumn{1}{l|}{89.06} & \multicolumn{1}{l|}{89.17} & 89.16 & \multicolumn{1}{l|}{94.72} & \multicolumn{1}{l|}{94.32} & \multicolumn{1}{l|}{95.13} & 94.72 & \multicolumn{1}{l|}{91.94} & \multicolumn{1}{l|}{88.84} & \multicolumn{1}{l|}{92.13} & 90.45 & \multicolumn{1}{l|}{84.64} & \multicolumn{1}{l|}{83.76} & \multicolumn{1}{l|}{85.18} & 84.46 \\ \hline
      External Memory-w/o & \multicolumn{1}{l|}{86.98} & \multicolumn{1}{l|}{86.98} & \multicolumn{1}{l|}{87.31} & 87.14 & \multicolumn{1}{l|}{93.95} & \multicolumn{1}{l|}{93.48} & \multicolumn{1}{l|}{92.26} & 92.86 & \multicolumn{1}{l|}{89.38} & \multicolumn{1}{l|}{86.83} & \multicolumn{1}{l|}{88.46} & 87.64 & \multicolumn{1}{l|}{84.11} & \multicolumn{1}{l|}{82.88} & \multicolumn{1}{l|}{84.21} & 83.53 \\ \hline
      Multistage Fusion-w/o & \multicolumn{1}{l|}{84.90} & \multicolumn{1}{l|}{84.9} & \multicolumn{1}{l|}{84.79} & 84.84 & \multicolumn{1}{l|}{91.67} & \multicolumn{1}{l|}{90.13} & \multicolumn{1}{l|}{90.41} & 90.26 & \multicolumn{1}{l|}{87.28} & \multicolumn{1}{l|}{84.69} & \multicolumn{1}{l|}{85.29} & 84.98 & \multicolumn{1}{l|}{81.12} & \multicolumn{1}{l|}{80.03} & \multicolumn{1}{l|}{82.11} & 81.05 \\ \hline
      \end{tabular}
  }
  \arrayrulecolor{black}
  \end{table}

\par It is noteworthy that in the absence of feature interaction, the F1 scores of the student model exhibited a decline of
 4.68\%, 3.45\%, 8.45\%, and 4.43\% for the four vulnerabilities. The multi-stage convolution fusion mechanism had the most 
 significant impact on the model's training outcomes, with an average reduction in accuracy of 5.25\% across all 
 vulnerabilities. From a technical perspective, during training, the model's top layers will accumulate a large amount of 
 semantic feature information. The multi-stage convolution fusion enhancement mechanism compensates for the deficiencies 
 in feature representation under complex data conditions, ensuring the effective integration of local features with global 
 information. Without this mechanism, the model may lose important semantic features, resulting in reduced fusion 
 capability and, consequently, decreased overall performance. Further analysis of the results in the table reveals that 
 the impact on model performance of removing the multi-dimensional query mechanism is smaller than that of removing the 
 external memory modeling mechanism. This is because the core task of model training is to extract fine-grained local 
 features through the multi-dimensional convolution fusion mechanism, and gradually enhance the relationship between local 
 and global features using the external memory space. Therefore, even without this mechanism, other modules can still 
 effectively ensure local feature extraction, leading to a relatively limited performance drop.

 \subsubsection{Knowledge Distillation Performance Assessment}
\begin{figure}[!ht]
   \centering
   \includegraphics[width=0.8\textwidth]{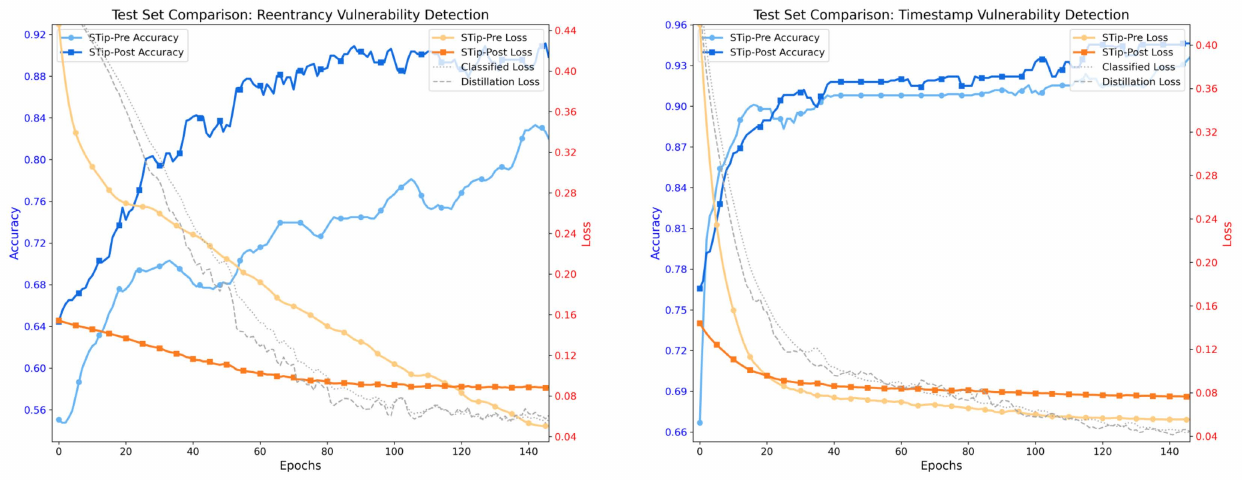}
   \caption{Comparison of STip model loss and accuracy before and after data-free knowledge distillation (Reentrancy vulnerability and Timestamp vulnerability test set)}
   \label{fig5}
 \end{figure}

\begin{figure}[!ht]
  \centering
  \includegraphics[width=0.8\textwidth]{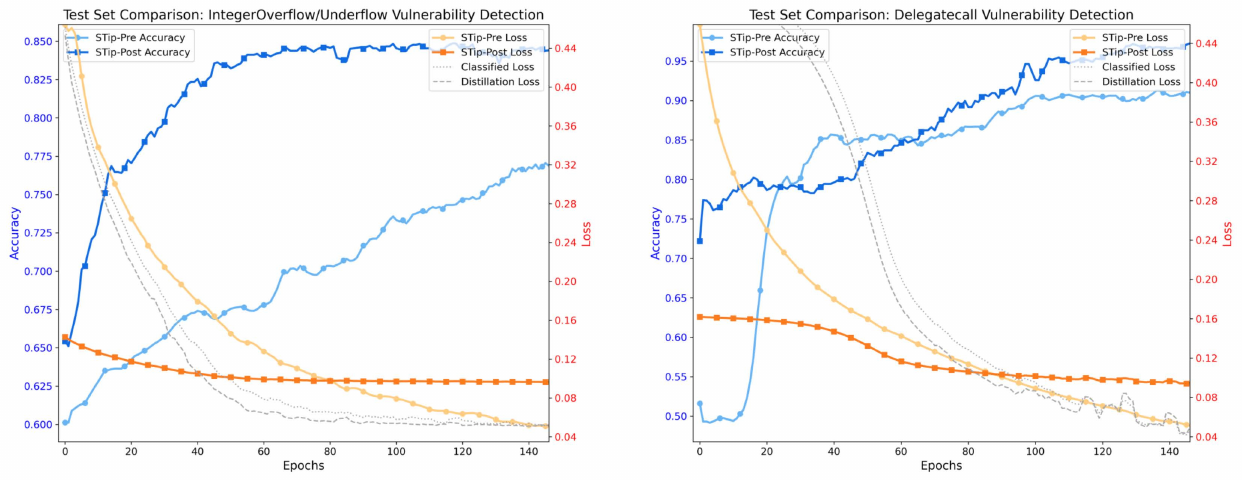}
  \caption{Comparison of STip model loss and accuracy before and after data-free knowledge distillation (IntegerOverflow/Underflow vulnerability and Delegatecall vulnerability test set)}
  \label{fig6}
\end{figure}

 In this section, we present the performance of the model on the four vulnerability test sets using graphs
  (Figures \ref{fig5}-\ref{fig6}). 
 Since the total loss of the distilled student model is calculated as a weighted sum of the classification loss and the 
 distillation loss, it typically shows smaller and more stable fluctuations. In contrast, the pre-distillation model relies 
 solely on traditional hard-label training, resulting in larger loss differences and stronger fluctuations, making it
  difficult to effectively compare the total loss of the pre-distillation model with that of the distilled model on the 
  same graph. In order to visually present the loss differences before and after distillation, and to ensure comparability on 
 a consistent scale, we applied min-max scaling to map the pre-distillation loss values to the range [0.04, 0.5].

\par As illustrated in the figure, the model's performance on the complex dataset demonstrates a 
notable enhancement with the increase in training epochs. In the reentrancy and integer overflow vulnerability test 
sets, the accuracy of STip (Post) exhibits a discernible superiority over that of STip (Pre). Furthermore, the 
loss area of the distilled model exhibits a substantial reduction, indicating that distillation enhances the model's 
stability and accuracy, with accuracy stabilising around 90\%. In contrast, the timestamp and delegatecall 
vulnerability test sets show smaller fluctuations in the loss curves before and after distillation, with the 
change in the loss area being relatively smooth. However, a deeper analysis of the data curves reveals that the 
loss area of the distilled model still decreases significantly. This indicates that for more complex datasets, such 
as the Delegatecall and integer overflow vulnerabilities, the proposed data-free distillation method enables the model
 to effectively enhance its learning and generalization 
capabilities even without the original data. This validates the model's advantages in handling complex tasks.

\subsubsection{Model Generalisation Performance}
\begin{figure}[!ht]
  \centering
  \includegraphics[width=1\textwidth]{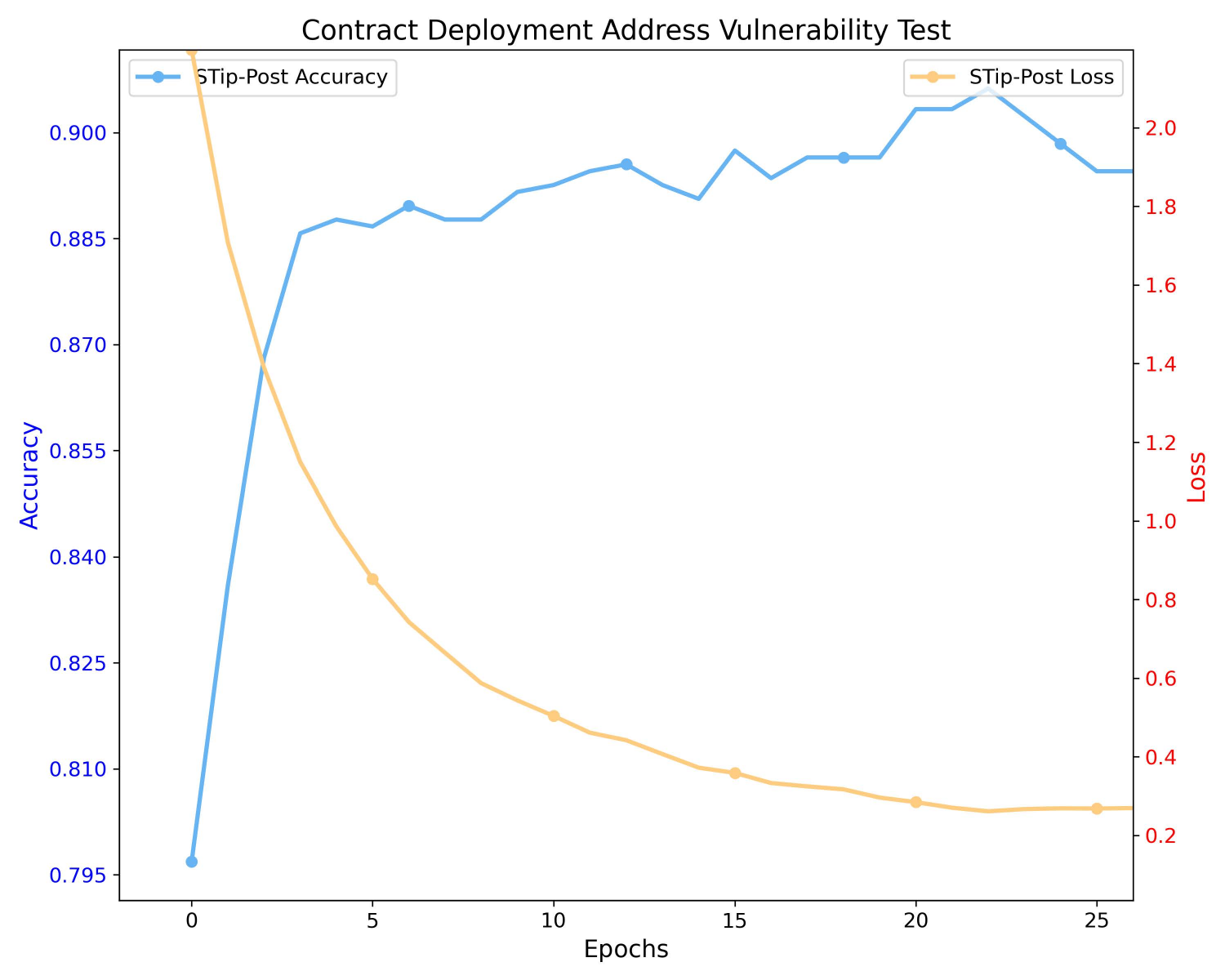}
  \caption{AF-STip model migration learning results after no data distillation (new vulnerability type test set)}
  \label{fig7}
\end{figure}

To validate the scalability of the STip model following data-free distillation, we elected to construct a novel 
vulnerability pertaining to contract deployment address types and conducted performance testing through transfer 
learning. Specifically, the student model was randomly selected after data-free distillation and its complete weight 
information was loaded. The same training parameters that were used before distillation were then applied to conduct 
30 epochs of transfer learning on the new 
vulnerability type dataset. The results are visualised in Figure \ref{fig7}, with the training parameters provided in Section
 4.2.

\section{Conclusion}
In this paper, we propose a data-free distillation-based multi-scale attention fusion model for smart contract
 vulnerability detection. The aim of our research is to explore how to improve the generalization ability of models in 
 the field of smart contract detection without additional training data (i.e., the lack of extra smart contract source
  code). In the data preprocessing stage, we employ word embedding techniques to capture the semantic information of 
  smart contract source code, ensuring the model can understand the semantic details within the contract. In the
   feature extraction stage, we introduce a novel multi-scale attention fusion module that performs bottom-up, 
   stage-wise feature interaction extraction, optimizing the model's ability to model key features. In the 
   classification stage, we use the teacher network as the main network and the student network as the target network, 
   adopting a data-free distillation strategy. The teacher model guides the student model with its knowledge, enabling 
   it to learn 
efficiently even in the absence of a large labeled dataset, thereby enhancing the student model's classification 
capability.

\par After experimental validation, the model proposed in this paper has higher accuracy and classification ability 
compared to other mainstream vulnerability detection models when targeting the four common vulnerabilities of current 
smart contracts. Furthermore, the utilisation of data-free knowledge distillation Even in the absence of additional 
training data, our lightly designed student model demonstrates a robust capacity for generalization. We have similarly 
elected to disseminate our dataset and the methodology underlying its construction. It encompasses sample processing 
data for five common vulnerabilities and delineates integer overflow vulnerabilities into upper and lower categories, 
with the objective of further advancing the field of smart contract vulnerability detection. We posit that as more 
researchers disseminate data and technology, 
blockchain smart contract vulnerability detection technology will evolve towards more efficient and accurate solutions.

\section*{Statement}
This work has been submitted to the IEEE for possible publication. Copyright may be transferred without notice, after which this version may no longer be accessible.
\end{CJK}
\bibliographystyle{unsrt}  
\bibliography{References} 

\begin{thebibliography}{10}

\bibitem{mohanta2018overview}
Bhabendu~Kumar Mohanta, Soumyashree~S Panda, and Debasish Jena.
\newblock An overview of smart contract and use cases in blockchain technology.
\newblock In {\em 2018 9th international conference on computing, communication
  and networking technologies (ICCCNT)}, pages 1--4. IEEE, 2018.

\bibitem{inspex2023curveattack}
Inspex.
\newblock Challenges and risk analysis in defi: The curve attack.
\newblock
  \url{https://medium.com/spoolfi/challenges-and-risk-analysis-in-defi-the-curve-attack-568d49fbe0b7},
  2023.
\newblock Accessed: 2024-12-15.

\bibitem{certik2023web3security}
CertiK.
\newblock Hack3d: The web3 security report 2023.
\newblock
  \url{https://www.certik.com/zh-CN/resources/blog/hack3d-the-web3-security-report-2023},
  2023.
\newblock Accessed: 2024-12-15.

\bibitem{shen2023smart}
Xueli Shen and Mingfeng Li.
\newblock Smart contract reentrancy vulnerability detection method based on
  deep learning hybrid model.
\newblock In {\em 2023 5th International Conference on Artificial Intelligence
  and Computer Applications (ICAICA)}, pages 33--36. IEEE, 2023.

\bibitem{gogineni2020multi}
Ajay~K Gogineni, Soumya Swayamjyoti, Devadatta Sahoo, Kisor~K Sahu, and Raj
  Kishore.
\newblock Multi-class classification of vulnerabilities in smart contracts
  using awd-lstm, with pre-trained encoder inspired from natural language
  processing.
\newblock {\em IOP SciNotes}, 1(3):035002, 2020.

\bibitem{kalra2018zeus}
Sukrit Kalra, Seep Goel, Mohan Dhawan, and Subodh Sharma.
\newblock Zeus: analyzing safety of smart contracts.
\newblock In {\em Ndss}, pages 1--12, 2018.

\bibitem{hinton2015distilling}
Geoffrey Hinton.
\newblock Distilling the knowledge in a neural network.
\newblock {\em arXiv preprint arXiv:1503.02531}, 2015.

\bibitem{wei2024dynamic}
Yukang Wei and Yu~Bai.
\newblock Dynamic temperature knowledge distillation.
\newblock {\em arXiv preprint arXiv:2404.12711}, 2024.

\bibitem{long2024mkdat}
Jun Long, Zhuoying Yin, Yan Han, and Wenti Huang.
\newblock Mkdat: Multi-level knowledge distillation with adaptive temperature
  for distantly supervised relation extraction.
\newblock {\em Information}, 15(7):382, 2024.

\bibitem{zagoruyko2016paying}
Sergey Zagoruyko and Nikos Komodakis.
\newblock Paying more attention to attention: Improving the performance of
  convolutional neural networks via attention transfer.
\newblock {\em arXiv preprint arXiv:1612.03928}, 2016.

\bibitem{furlanello2018born}
Tommaso Furlanello, Zachary Lipton, Michael Tschannen, Laurent Itti, and Anima
  Anandkumar.
\newblock Born again neural networks.
\newblock In {\em International conference on machine learning}, pages
  1607--1616. PMLR, 2018.

\bibitem{wang2023improving}
Yuzhu Wang, Lechao Cheng, Manni Duan, Yongheng Wang, Zunlei Feng, and Shu Kong.
\newblock Improving knowledge distillation via regularizing feature norm and
  direction.
\newblock {\em arXiv preprint arXiv:2305.17007}, 2023.

\bibitem{park2019relational}
Wonpyo Park, Dongju Kim, Yan Lu, and Minsu Cho.
\newblock Relational knowledge distillation.
\newblock In {\em Proceedings of the IEEE/CVF conference on computer vision and
  pattern recognition}, pages 3967--3976, 2019.

\bibitem{yang2022cross}
Chenxiao Yang, Junwei Pan, Xiaofeng Gao, Tingyu Jiang, Dapeng Liu, and Guihai
  Chen.
\newblock Cross-task knowledge distillation in multi-task recommendation.
\newblock In {\em Proceedings of the AAAI conference on artificial
  intelligence}, volume~36, pages 4318--4326, 2022.

\bibitem{shao2022review}
Renrong Shao, Yu'ang Liu, Wei Zhang, and Jun Wang.
\newblock A survey of knowledge distillation in deep learning.
\newblock {\em Journal of Computer Science (Jisuanji Xuebao)},
  45(8):1638--1673, 2022.

\bibitem{vidal2024vulnerability}
Fernando~Richter Vidal, Naghmeh Ivaki, and Nuno Laranjeiro.
\newblock Vulnerability detection techniques for smart contracts: A systematic
  literature review.
\newblock {\em Journal of Systems and Software}, page 112160, 2024.

\bibitem{luu2016making}
Loi Luu, Duc-Hiep Chu, Hrishi Olickel, Prateek Saxena, and Aquinas Hobor.
\newblock Making smart contracts smarter.
\newblock In {\em Proceedings of the 2016 ACM SIGSAC conference on computer and
  communications security}, pages 254--269, 2016.

\bibitem{ruskin1980mythril}
Laura Ruskin.
\newblock Mythril\# 8.
\newblock {\em Mythril}, 2(4):1, 1980.

\bibitem{tsankov2018securify}
Petar Tsankov, Andrei Dan, Dana Drachsler-Cohen, Arthur Gervais, Florian
  Buenzli, and Martin Vechev.
\newblock Securify: Practical security analysis of smart contracts.
\newblock In {\em Proceedings of the 2018 ACM SIGSAC conference on computer and
  communications security}, pages 67--82, 2018.

\bibitem{mossberg2019manticore}
Mark Mossberg, Felipe Manzano, Eric Hennenfent, Alex Groce, Gustavo Grieco,
  Josselin Feist, Trent Brunson, and Artem Dinaburg.
\newblock Manticore: A user-friendly symbolic execution framework for binaries
  and smart contracts.
\newblock In {\em 2019 34th IEEE/ACM International Conference on Automated
  Software Engineering (ASE)}, pages 1186--1189. IEEE, 2019.

\bibitem{zhang2022cbgru}
Lejun Zhang, Weijie Chen, Weizheng Wang, Zilong Jin, Chunhui Zhao, Zhennao Cai,
  and Huiling Chen.
\newblock Cbgru: A detection method of smart contract vulnerability based on a
  hybrid model.
\newblock {\em Sensors}, 22(9):3577, 2022.

\bibitem{liu2023smart}
Zhenpeng Liu, Mingxiao Jiang, Shengcong Zhang, Jialiang Zhang, and Yi~Liu.
\newblock A smart contract vulnerability detection mechanism based on deep
  learning and expert rules.
\newblock {\em IEEE Access}, 2023.

\bibitem{zhang2022novel}
Lejun Zhang, Jinlong Wang, Weizheng Wang, Zilong Jin, Chunhui Zhao, Zhennao
  Cai, and Huiling Chen.
\newblock A novel smart contract vulnerability detection method based on
  information graph and ensemble learning.
\newblock {\em Sensors}, 22(9):3581, 2022.

\bibitem{zhang2022spcbig}
Lejun Zhang, Yuan Li, Tianxing Jin, Weizheng Wang, Zilong Jin, Chunhui Zhao,
  Zhennao Cai, and Huiling Chen.
\newblock Spcbig-ec: a robust serial hybrid model for smart contract
  vulnerability detection.
\newblock {\em Sensors}, 22(12):4621, 2022.

\bibitem{liu2021combining}
Zhenguang Liu, Peng Qian, Xiaoyang Wang, Yuan Zhuang, Lin Qiu, and Xun Wang.
\newblock Combining graph neural networks with expert knowledge for smart
  contract vulnerability detection.
\newblock {\em IEEE Transactions on Knowledge and Data Engineering},
  35(2):1296--1310, 2021.

\bibitem{liu2021smart}
Zhenguang Liu, Peng Qian, Xiang Wang, Lei Zhu, Qinming He, and Shouling Ji.
\newblock Smart contract vulnerability detection: from pure neural network to
  interpretable graph feature and expert pattern fusion.
\newblock {\em arXiv preprint arXiv:2106.09282}, 2021.

\bibitem{wu2021peculiar}
Hongjun Wu, Zhuo Zhang, Shangwen Wang, Yan Lei, Bo~Lin, Yihao Qin, Haoyu Zhang,
  and Xiaoguang Mao.
\newblock Peculiar: Smart contract vulnerability detection based on crucial
  data flow graph and pre-training techniques.
\newblock In {\em 2021 IEEE 32nd International Symposium on Software
  Reliability Engineering (ISSRE)}, pages 378--389. IEEE, 2021.

\bibitem{chen2023smart}
Da~Chen, Lin Feng, Yuqi Fan, Siyuan Shang, and Zhenchun Wei.
\newblock Smart contract vulnerability detection based on semantic graph and
  residual graph convolutional networks with edge attention.
\newblock {\em Journal of Systems and Software}, 202:111705, 2023.

\bibitem{li2018vuldeepecker}
Zhen Li, Deqing Zou, Shouhuai Xu, Xinyu Ou, Hai Jin, Sujuan Wang, Zhijun Deng,
  and Yuyi Zhong.
\newblock Vuldeepecker: A deep learning-based system for vulnerability
  detection.
\newblock {\em arXiv preprint arXiv:1801.01681}, 2018.

\bibitem{yue2020sentiment}
Wang Yue and Lei Li.
\newblock Sentiment analysis using word2vec-cnn-bilstm classification.
\newblock In {\em 2020 seventh international conference on social networks
  analysis, management and security (SNAMS)}, pages 1--5. IEEE, 2020.

\bibitem{qian2023cross}
Peng Qian, Zhenguang Liu, Yifang Yin, and Qinming He.
\newblock Cross-modality mutual learning for enhancing smart contract
  vulnerability detection on bytecode.
\newblock In {\em Proceedings of the ACM Web Conference 2023}, pages
  2220--2229, 2023.

\bibitem{mikolov2013distributed}
Tomas Mikolov, Ilya Sutskever, Kai Chen, Greg~S Corrado, and Jeff Dean.
\newblock Distributed representations of words and phrases and their
  compositionality.
\newblock {\em Advances in neural information processing systems}, 26, 2013.

\bibitem{bojanowski2017enriching}
Piotr Bojanowski, Edouard Grave, Armand Joulin, and Tomas Mikolov.
\newblock Enriching word vectors with subword information.
\newblock {\em Transactions of the association for computational linguistics},
  5:135--146, 2017.

\bibitem{vaswani2017attention}
A~Vaswani.
\newblock Attention is all you need.
\newblock {\em Advances in Neural Information Processing Systems}, 2017.

\bibitem{zhang2022epsanet}
Hu~Zhang, Keke Zu, Jian Lu, Yuru Zou, and Deyu Meng.
\newblock Epsanet: An efficient pyramid squeeze attention block on
  convolutional neural network.
\newblock In {\em Proceedings of the asian conference on computer vision},
  pages 1161--1177, 2022.

\bibitem{dai2021coatnet}
Zihang Dai, Hanxiao Liu, Quoc~V Le, and Mingxing Tan.
\newblock Coatnet: Marrying convolution and attention for all data sizes.
\newblock {\em Advances in neural information processing systems},
  34:3965--3977, 2021.

\bibitem{durieux2020empirical}
Thomas Durieux, Jo{\~a}o~F Ferreira, Rui Abreu, and Pedro Cruz.
\newblock Empirical review of automated analysis tools on 47,587 ethereum smart
  contracts.
\newblock In {\em Proceedings of the ACM/IEEE 42nd International conference on
  software engineering}, pages 530--541, 2020.

\bibitem{nguyen2020sfuzz}
Tai~D Nguyen, Long~H Pham, Jun Sun, Yun Lin, and Quang~Tran Minh.
\newblock sfuzz: An efficient adaptive fuzzer for solidity smart contracts.
\newblock In {\em Proceedings of the ACM/IEEE 42nd International Conference on
  Software Engineering}, pages 778--788, 2020.

\bibitem{torres2018osiris}
Christof~Ferreira Torres, Julian Sch{\"u}tte, and Radu State.
\newblock Osiris: Hunting for integer bugs in ethereum smart contracts.
\newblock In {\em Proceedings of the 34th annual computer security applications
  conference}, pages 664--676, 2018.

\bibitem{tikhomirov2018smartcheck}
Sergei Tikhomirov, Ekaterina Voskresenskaya, Ivan Ivanitskiy, Ramil Takhaviev,
  Evgeny Marchenko, and Yaroslav Alexandrov.
\newblock Smartcheck: Static analysis of ethereum smart contracts.
\newblock In {\em Proceedings of the 1st international workshop on emerging
  trends in software engineering for blockchain}, pages 9--16, 2018.

\bibitem{feist2019slither}
Josselin Feist, Gustavo Grieco, and Alex Groce.
\newblock Slither: a static analysis framework for smart contracts.
\newblock In {\em 2019 IEEE/ACM 2nd International Workshop on Emerging Trends
  in Software Engineering for Blockchain (WETSEB)}, pages 8--15. IEEE, 2019.

\bibitem{tann2018towards}
Wesley Joon-Wie Tann, Xing~Jie Han, Sourav~Sen Gupta, and Yew-Soon Ong.
\newblock Towards safer smart contracts: A sequence learning approach to
  detecting security threats.
\newblock {\em arXiv preprint arXiv:1811.06632}, 2018.

\bibitem{qian2020towards}
Peng Qian, Zhenguang Liu, Qinming He, Roger Zimmermann, and Xun Wang.
\newblock Towards automated reentrancy detection for smart contracts based on
  sequential models.
\newblock {\em IEEE Access}, 8:19685--19695, 2020.

\bibitem{kipf2016semi}
Thomas~N Kipf and Max Welling.
\newblock Semi-supervised classification with graph convolutional networks.
\newblock {\em arXiv preprint arXiv:1609.02907}, 2016.

\end{thebibliography}

\end{document}